\documentclass{aa}

\usepackage{graphicx}

\newbox\grsign \setbox\grsign=\hbox{$>$} \newdimen\grdimen \grdimen=\ht\grsign
\newbox\simlessbox \newbox\simgreatbox
\setbox\simgreatbox=\hbox{\raise.5ex\hbox{$>$}\llap
     {\lower.5ex\hbox{$\sim$}}}\ht1=\grdimen\dp1=0pt
\setbox\simlessbox=\hbox{\raise.5ex\hbox{$<$}\llap
     {\lower.5ex\hbox{$\sim$}}}\ht2=\grdimen\dp2=0pt

\newbox\simppropto
\setbox\simppropto=\hbox{\raise.5ex\hbox{$\sim$}\llap
     {\lower.5ex\hbox{$\propto$}}}\ht2=\grdimen\dp2=0pt

\begin{document}

\title{Gemini-Phoenix infrared 
high-resolution abundance analysis of five giants
in the bulge globular cluster NGC 6553
\thanks{Observations collected at the
Gemini Observatory, Chile, and the European Southern Observatory,
 Paranal, Chile} }

\author{J. Mel\'endez\inst{1,2}, B. Barbuy\inst{1}, E. Bica\inst{3},
 M. Zoccali\inst{4}, S. Ortolani\inst{5}, A. Renzini\inst{4},
V. Hill\inst{6}}

\institute{
Universidade de S\~ao Paulo, Depto. de Astronomia,  
Rua do Mat\~ao 1226, Cid. Universit\'aria, 
S\~ao Paulo 05508-900, Brazil.
e-mail: jorge@astro.iag.usp.br, barbuy@astro.iag.usp.br
\and
Universidad Nacional Mayor de San Marcos, Seminario Permanente de Astronom\'{\i}a y
Ciencias Espaciales, Facultad de Ciencias F\'{\i}sicas, Av. Venezuela s/n, Lima 1, Peru.
\and
Universidade Federal do Rio Grande do Sul, Dept. de Astronomia, 
CP 15051, Porto Alegre 91501-970, Brazil.
e-mail: bica@if.ufrgs.br
\and
European Southern Observatory, Karl-Schwarzschild Strasse 2, D-85748, 
Garching bei M\"unchen, Germany. e-mail: mzoccali@eso.org, arenzini@eso.org
\and
Universit\`a di Padova, Dipartimento di Astronomia, Vicolo
 dell'Osservatorio 5, I-35122 Padova, Italy.
e-mail: ortolani@pd.astro.it
\and
Observatoire de Paris-Meudon, 92195 Meudon Cedex, France.
e-mail: vanessa.hill@obspm.fr
}

\date{Received ; accepted }

\abstract{
 A detailed abundance analysis of 5 giants of the metal-rich
bulge globular cluster NGC 6553 was  carried out using 
high resolution infrared spectra in the H band,
 obtained at the Gemini-South 8m telescope.\\
 JK photometry collected at ESO
and VI photometry from the Hubble Space Telescope
are used to derive effective temperatures.
The present analysis provides a metallicity
[Fe/H] = -0.20$\pm$0.10. An overabundance of oxygen of [O/Fe] = +0.20
is found from IR OH lines. 
}

\titlerunning{IR analysis of giants in NGC 6553}

\authorrunning{J. Mel\'endez et al.} 

\maketitle

\section{Introduction.} 

NGC 6553 is a metal-rich globular cluster,
projected in the direction of the Galactic bulge,
 located at $\alpha_{2000}$ = 
18$^{\rm h}$09$^{\rm m}$16$^{\rm s}$, $\delta_{2000}$ =
 -25$^{\circ}$54'28",
and (l,b) = (5.25$^{\circ}$,-3.02$^{\circ}$)
 at a distance of d$_{\odot}$ $\approx$ 5.1 kpc
(Guarnieri et al. 1998). Ortolani et al. (1990, 1995)
have shown the turnover of the Red Giant Branch
(RGB), due to the presence of TiO bands,
in Colour-Magnitude Diagram (CMD) of NGC 6553. 
This RGB morphology indicates that NGC 6553
is more metal-rich than the halo cluster 47 Tucanae,
and almost identical to  the bulge cluster NGC 6528
(Ortolani et al. 1995;  
Momany et al. 2003). 
NGC 6553, together with NGC 6528, are
also shown to have a  metallicity and an  age comparable
to the bulk of the Baade`s Window  stellar population
(Ortolani et al. 1995; Zoccali et al. 2003a).
Due to their high metallicities,
  NGC 6528, NGC 6553 and NGC 6440 were adopted by Bica (1988)
as templates 
 to reproduce the spectra of the more metal-rich elliptical 
galaxies.
 
NGC 6553, as a reference metal-rich cluster,
is suitable for studies of individual stars, because
it is relatively loose and uncrowded,
and shows a reddening considerably lower than the majority
of metal-rich globular clusters.

A proper motion study of NGC 6553 (Zoccali et al. 2001, 2003b) showed
that its  orbital
rotational velocity around the Galactic center
  is consistent with the mean rotation of the
 bulge at 2.7 kpc (Minniti 1995) and as well with the disk rotation
at the same distance (Amaral et al. 1996). 
Given that from the kinematical point of view, 
NGC 6553 appears to be consistent  with either a
disk or a bulge membership, it is important to 
 solve this  issue 
by deriving elemental abundance ratios,
as signatures from star formation timescale and age.

The analyses of individual stars of NGC 6553 have 
given however some conflicting results:

Barbuy et al. (1992) estimated [Fe/H] $\approx$ -0.2,
from fits of synthetic spectra to the observed
spectrum of a cool giant at a resolution of R $\sim$ 20 000; 
no abundance ratios were derived
given the limitations of S/N.
From higher signal-to-noise data, with the same resolution,
 Barbuy et al. (1999) derived [Fe/H] =
-0.55, together with significant $\alpha$-element enhancements
for two giants of effective temperatures around 4000 K.
Cohen et al. (1999) analysed 5 Horizontal Branch (HB) stars
and found a mean value of  [Fe/H] = -0.16, with
the mean overabundances of $\alpha$-elements
[Mg/Fe] = +0.4, [Si/Fe] = +0.14, [Ca/Fe] = +0.26,
[Ti/Fe] = +0.19, whereas the oxygen  value of
[O/Fe] = +0.5 may be overestimated given the use
of the OI 777nm triplet, which tends to give
higher values than other oxygen lines 
(e.g. Mel\'endez et al. 2001; Lambert 2002).

Carretta et al. (2001) analysed 4 HB stars of NGC 6528,
having obtained [Fe/H] = +0.07, and they  proposed a rescaling of
the metallicity of NGC 6553 to [Fe/H] = -0.06,
0.1dex higher than given in Cohen et al. (1999).

Origlia et al. (2002) analysed two cool giants of effective
temperature T$_{\rm eff}$ = 4000 K from infrared spectra
in the H band, at a resolution R $\sim$ 25 000,
 and found a mean metallicity of 
[Fe/H] = -0.3 and [$\alpha$/Fe] = +0.3.

In the present work we analyse 5 giants of NGC 6553,
using high resolution H-band spectra obtained with
the Phoenix spectrograph at the Gemini-South 8m telescope.
At a resolution R $\sim$ 50 000,
these are the highest resolution observations so far
obtained for stars in this cluster.
 
In Sect. 2 the observations are described. In Sect. 3 the stellar parameters
effective temperature, gravity, metallicity 
and abundance ratios are derived. In Sect. 4
the results are discussed. In Sect. 5 conclusions are drawn.

\begin{table*}
\caption[1]{Log of observations}
\begin{flushleft}
\begin{tabular}{lllrlllrrrlll}
\hline \hline
\noalign{\smallskip}
\noalign{\vskip 0.1cm}
{\rm Star} & {\rm V} & {\rm I$_{\rm C}$} &{\rm J}&{\rm K} &{\rm Date} & 
{\rm Exp. time} & {\rm S/N} & {\rm S/N} &{\rm v$^{\rm hel}_{\rm r}$} \cr
{\rm} &  &  & &  &{\rm } & 
{\rm (s)} & {\rm resol.} & {\rm pixel} &{\rm km s$^{-1}$} \cr
\noalign{\vskip 0.1cm}
\noalign{\hrule\vskip 0.1cm}
\noalign{\vskip 0.1cm}
\multicolumn{10}{c}{Gemini-South}\\
\noalign{\vskip 0.1cm}
\noalign{\hrule\vskip 0.1cm}
40201 (III-17)& 15.36 &12.35 &10.31 & 8.97 &2002 May 06 & 6 
$\times$ 420 &330 &200 &-8.7\cr
20150         & 15.40 &12.67 &10.81 & 9.45 &2002 May 08 & 3 
$\times$ 600 &240 &120 & 7.7\cr
40056 (II-85) & 15.52 &13.01 &11.04 & 9.79 &2002 May 09 & 9 
$\times$ 600 &240 &150 &1.6\cr 
20074 (IV-13) & 15.59 &13.11 &11.28 &10.03 &2002 May 08 & 9 
$\times$ 600 &260 &130 & 9.3\cr
40082 (III-3) & 15.82 &13.41 &11.54 &10.33 &2002 May 09 & 3 
$\times$ 600 &160  &80 &-1.0\cr
\noalign{\vskip 0.1cm}
\noalign{\hrule\vskip 0.1cm}
\multicolumn{10}{c}{VLT}\\
\noalign{\vskip 0.1cm}
\noalign{\hrule\vskip 0.1cm}
40056 (II-85) & 15.52 &13.01 &11.04 & 9.79 &2000 Jun 26-27 & 2 $\times$ 3600 & 170 &70 &0.5\cr 
\noalign{\smallskip} \hline \end{tabular}
\end{flushleft} 
\end{table*}

\section{Observations}

The log-book of observations is reported in Table 1.
The spectra of individual stars of
NGC 6553 in the H band, at a resolution of R = 50 000,
were obtained using the 8 m Gemini South telescope and
the Phoenix spectrograph (Hinkle et al. 2000).
The instrument uses a 1024 $\times$ 1024 InSb Aladdin array, which
is sensitive in the 1-5 $\mu$m region. The observations were
centred at 1.555 $\mu$m, with a wavelength coverage of 75 {\rm \AA}. The
spectra were observed using the 4-pixel slit ($0\farcs35$).

Each sample star was observed at three different positions separated
by 3" along the slit. Most exposures were taken with an integration 
time of 10 min at each position. The sky and dark background were
removed by subtracting exposures taken at different positions on 
the detector array. Each night 10 flat and 10 dark frames
were gathered. 
A hot star was observed to check for the presence of
 telluric lines, and 
the region was found to be almost free of them.

The frames were reduced with IRAF, following a procedure similar
to that described in Mel\'endez et al. (2001). 
Dark and flat frames are combined
and the resulting dark is subtracted from the flat. A bad pixel mask is
obtained using the subtracted flat, and the bad pixels of the image frames
were replaced by interpolation of surrounding pixels. 
A response image was obtained from the 
corrected flat
and the sky corrected object frames were divided by the response frame. The
spectra were extracted and wavelength calibrated using the absorption 
stellar lines, and finally the spectra taken at different positions were
combined and the continuum was normalized.

The star II-85 was also observed in the optical,
 in the wavelength range
$\lambda\lambda$ 4800-6800 {\rm \AA} 
with the UVES spectrograph at the VLT UT2 telescope
at ESO (Paranal).
With the standard configuration,
a resolution of R $\sim$ 60 000 was achieved for a slit
width of $0.8''$.
In Table 1 the S/N values per pixel and per resolution
element are given.
The resolution element at the UVES spectrograph cover
7-pixels, with 0.017 ${\rm \AA}$ per pixel. 
The spectra were reduced  using the UVES context of the
MIDAS reduction package, including bias and
inter-order background subtraction,
flatfield correction, optimal extraction (with cosmic
ray rejection) above the sky level and wavelength calibration
(Balester et al. 2000). 

The targets are identified in Fig. 1, which is a deep
Hubble Space Telescope (HST) F814W image.

In Fig. 2 is shown the V vs. $V-I$ Colour-Magnitude
 Diagram of NGC 6553
using  HST
(Ortolani et al. 1995) observations, where
the sample stars are represented by filled circles.
The CMD is corrected from differential reddening 
following method by Piotto et al. (1999) and
Zoccali et al. (2001).
The identifications given in Hartwick (\cite{hartwick}) 
are adopted along
the text.

\begin{figure*} 
\includegraphics[bb=15 205 573 585,width=17cm]{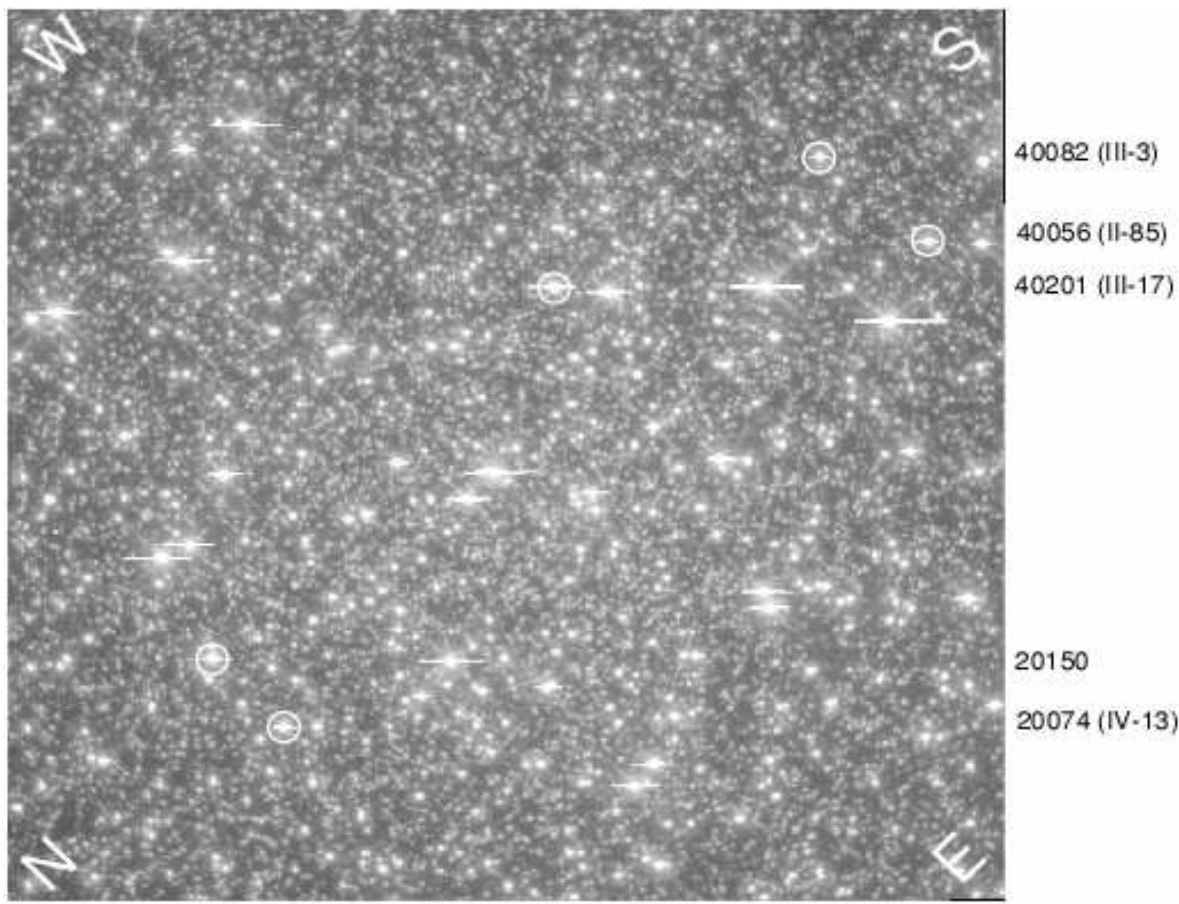}
\caption[]{Identification of the observed stars
in a F814W  field observed with the Hubble Space Telescope.
The field is about 1.3'$\times$1.3'.   The center of this map
essentially coincides with the center of NGC 6553.
}
\label{fig1}
\end{figure*}

The radial velocities, indicated in Table 1,
were derived with respect to the sky lines
reported in Table 2. 
A mean value of  v$^{\rm hel}_{\rm r}$  = 1.6$\pm$6 km s$^{-1}$ is obtained.
Fig. 3 shows a histogram of radial velocities
for the present values combined to stars from Origlia et
al. (2002), Barbuy et al. (1999) and Cohen et al. (1999). 

A peak value of v$^{\rm hel}_{\rm r}$  = 2 km s$^{-1}$  is found (Fig. 3), 
in good agreement with the value of 8.4$\pm$8.4 kms$^{-1}$
   measured by Rutledge et al. (1997).
It is also consistent with the mean value derived,
from low resolution spectra
using three methods, by Coelho et al. (2001), of  
v$^{\rm hel}_{\rm r}$  $\approx$ -1 km s$^{-1}$; 
membership of most sample 
stars of NGC 6553 were examined in Coelho et al. (2001) 
 prior to observations with the large telescopes.

\begin{table}
\caption[]{OH sky lines}
\begin{flushleft}
\begin{tabular}{lllllllllllll}
\hline
\noalign{\smallskip}
{\rm $\lambda$ (\AA)} & Transition & Wavenumber (cm$^{-1})$ & Int. \\
\noalign{\smallskip}
\hline
\noalign{\smallskip}
    15535.462 & (3, 1) P1e 5.5   & 6435.128 &  1.0 \\
    15536.705 & (3, 1) P1f 5.5   & 6434.613 &  1.0 \\
    15541.645 & (4, 2) R1f 3.5   & 6432.568 &  0.7 \\
    15542.146 & (4, 2) R1e 3.5   & 6432.360 &  0.7 \\
    15565.907*& (4, 2) R2e,f 3.5 & 6422.542 &  0.8 \\
\noalign{\smallskip} \hline \end{tabular}
\renewcommand{\arraystretch}{1}
\begin{list}{}{}
\item [$^{*}$] Blend:  15565.817 \& 15565.996 \AA
\end{list}
\end{flushleft} 
\end{table}

\begin{figure} 
\includegraphics[width=9cm]{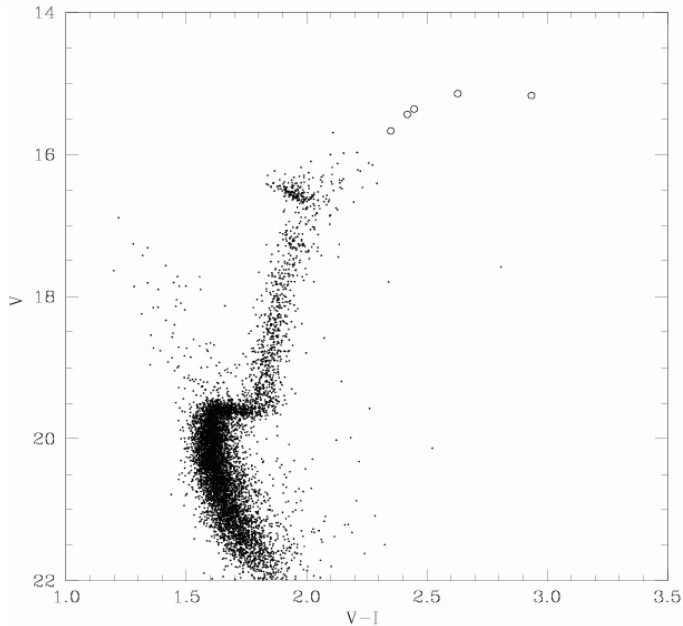}
\caption[]{V vs. V-I CMD of NGC 6553 using data from HST. 
Open circles: present sample of stars.
}
\label{fig2}
\end{figure}

\begin{figure} 
\includegraphics[width=9cm]{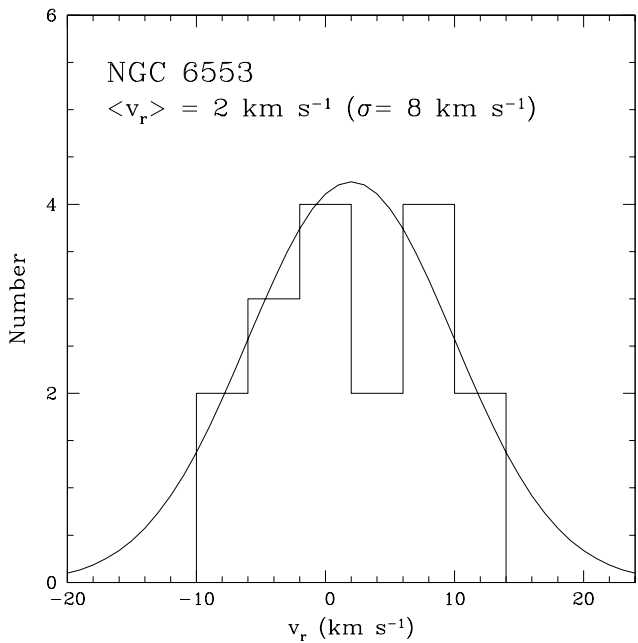}
\caption[]{Histogram of radial velocities using data from this work,
Origlia et al. (2002), Barbuy et al. (1999) and Cohen et al. (1999).
}
\label{fig3}
\end{figure}

\begin{figure} 
\includegraphics[width=15cm]{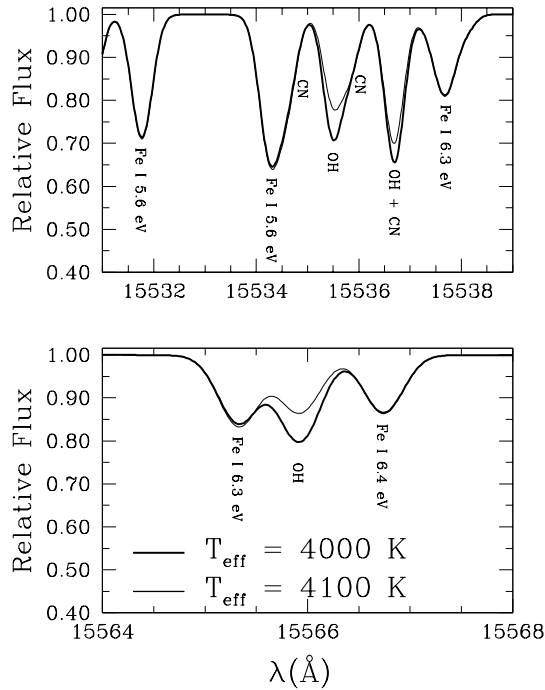}
\caption[]{Synthetic spectra with two different T$_{\rm eff}$,
showing that infrared (6 eV) iron lines are not sensitive to
a variation of $\Delta$ T$_{\rm eff}$ = +100 K.
}
\label{fig4}
\end{figure}

\begin{figure} 
\includegraphics[width=8cm]{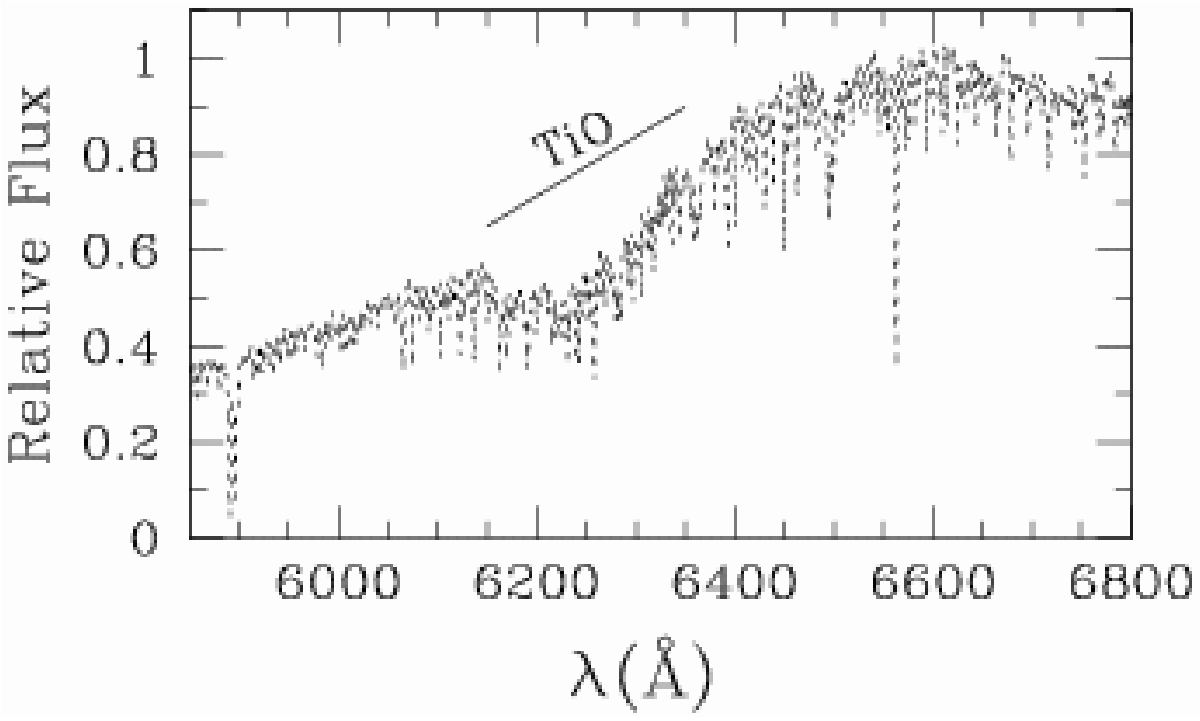}
\caption[]{VLT spectrum of star II-85, convolved with a gaussian
of sigma = 30 pixels, illustrating the region  affected 
by  TiO bands.
}
\label{figd}
\end{figure}

\section{Stellar Parameters}

\subsection{Reddening}

Zinn (1980) used integrated photometry in the
Gunn uvgr and two passbands on the H and K
CaII lines, and found E(B-V) = 0.78.
Reed et al. (1988) used integrated UBVRI colours
and estimated E(B-V) = 0.82.

Schlegel et al. (1998) presented a 100$\mu$m map of 
the Galaxy allowing to extract the reddening at different
galactic coordinates, corresponding to the total value
in each line of sight across the Galaxy, in the direction
of extragalactic sources. 
For $|$b$|$ $<$ 5$^{\circ}$ the
values are less precise. For NGC 6553, E(B-V) = 1.42 
is derived; Schlegel's extinctions are
overestimated in the bulge direction, and a
quantification of this is given by Dutra et al. (2003)
who showed that at these inner
Galaxy directions Schegel's values should be
multiplied by 0.75, giving thus E(B-V) = 1.06.

Cohen et al. (1999) analysed 5 HB stars, and estimated
their temperatures from excitation equilibrium of 
\ion{Fe}{I} lines. In order to be compatible with the colours,
they deduced E(B-V) = 0.78 for 3 stars and 0.86 for the two
other ones.

 Methods using CMDs
compare  the locus of the Red Giant Branch (RGB)
 of the cluster
with that of a cluster of similar metallicity,
or an isochrone.
Sagar et al. (1999)  compared
the CMD of NGC 6553 to isochrones and obtained
E(B-V) = 0.7.
 A comparison with 47 Tuc
([Fe/H] = -0.71, Harris 1996) was carried out by
Ortolani et al. (1990), Guarnieri et al. (1998) and Zoccali 
et al. (2001); the latter two papers used the same
data set and values E(V-I) = 0.95 and 0.80,
or E(B-V) = 0.7 and 0.6, respectively, were found.
The difference between these two values are due to two
reasons: 
(a) the colour of the RGB at the HB level was
given with a  $\Delta$(V-I) = 0.05 difference
between Zoccali et al. (2001) and
Guarnieri et al. (1998);
(b) a mistake in the derivation of Zoccali et al. (2001)
has been to consider that the reddening of 47 Tuc
had to be subtracted, whereas it has to be added,
giving thus a difference of  $\Delta$(V-I) = 0.1,
since E(V-I)$_{\rm 47 Tuc}$ = 0.05.
 
The blanketing
assumed for the difference between the metallicities of
47 Tuc and NGC 6553 is $\Delta(V-I)$ = 0.14 
(Girardi et al. 2002; Kim et al. 2002).
The V-I colours of the RGB at the HB level 
are (V-I)$_{\rm 47Tuc}$ = 1.0 (Barbuy et al. 1998)
 and (V-I)$_{\rm NGC 6553}$ = 2.08 (this refers
to the average value, as can be checked in Fig. 7 of
Zoccali et al. 2001).
Therefore we have: E(V-I)$_{\rm NGC 6553-47Tuc}$
= 2.08 - 1.0 - 0.14 + 0.05 = 0.99
 Assuming E(V-I)/E(B-V) = 1.33 (Dean et al. 1978),
or 1.36 (Schlegel et al. 1998),
this transforms to E(B-V) = 0.74 or 0.73 respectively.

A check of differential reddening for our sample stars,
following the method described in Zoccali et al. (2001),
indicates that they have a 
lower reddening, relative to the mean,
 of $\Delta$(B-V) = -0.04 for 3 stars, -0.055 for III-17
and -0.075 for 20150.  Therefore
in the present work, we assumed 
E(B-V) = 0.74 - 0.04 = 0.70 for all sample stars. 

\subsection{Effective Temperatures}
The  determination of effective temperatures T$_{\rm eff}$  was carried
out using V, I, J and K colours, reported in Table 1.  
V and I colours were obtained using HST
 and J and K colours using the detector IRAC2
at the ESO 2.2m telescope  (Ortolani et al. 1995; 
Guarnieri et al. 1998). 

Assuming a ratio 
E(V-I)/E(B-V)=1.33 (Dean et al. 1978),
E(V-K)/E(B-V) = 2.744 and E(J-K)/E(B-V) = 0.527
(Rieke \& Lebofsky 1985),  
the available colours were dereddened,
and they are given in Table 3.

In the present wavelength region the Fe I
lines have high excitation potentials around
5.5 - 6.5 eV, therefore the line strengths are not sensitive
to errors in temperature, as can be seen in Fig. 4. 
To constrain the effective temperature in a reddening-free
manner (excitation equilibrium), we also measured
  Fe I lines in a range of excitation potential values,
 from optical spectra of the star II-85. This star
has an effective temperature of T$_{\rm eff}$ $\approx$ 4000 K
and shows  TiO bandheads, as shown in Fig. 5.
We measured Fe I lines only in
the wavelength regions 5850 - 6150 {\rm \AA}
and 6360 - 6810 {\rm \AA}. In Fig. 6
are plotted the derived Fe abundances vs.
excitation potential $\chi_{\rm exc}$ and
reduced equivalent width.
It is clear that the T$_{\rm eff}$ = 4000 K is
in good agreement with the excitation 
equilibrium requirement. This indicates that the
effective temperatures derived from colours are
reliable, at least  for II-85. Full analysis of optical 
spectra obtained with the UVES spectrograph at the
Very Large Telescope is in progress (Hill et al., 
in preparation).

In Table 3 are reported effective temperatures
derived using the  $V-I$, $V-K$ and $J-K$ 
calibrations of Alonso, Arribas \& Mart\'{\i}nez-Roger (1999,  
hereafter AAM99) and those by Houdashelt et al. (2000a,b).
There is good agreement between these temperatures and derivations
based on relations by
Bessell et al. (1998) and  McWilliam (1990).

\begin{table*}
\caption[1]{Colours and derived effective temperatures.}
\begin{flushleft}
\begin{tabular}{clllllllllllllllll}
\hline \hline
\noalign{\smallskip}
Id. & No.&  E(B-V) & $V-I_{\rm C}$$_{0}$ & $V-K_{0}$ & $J-K_{0}$  & 
\multispan2 T$_{\rm eff}$(V-I) &   \multispan2 T$_{\rm eff}$(V-K) & 
 \multispan2 T$_{\rm eff}$(J-K)& \multispan2 Mean & \multispan2 Mean & Final \\
     &       &     &     &   &    &  Houd & Alon& Houd& Alon& Houd& Alon&    Houd& Alonso&  & \\
\noalign{\smallskip}
\hline
\noalign{\smallskip}
III17 & 40201 &0.70 & 2.06 & 4.51 & 0.87 & 3694& 3596& 3684& 3634 &3994& 3986 & 3790 & 3739 &3765 \\
      & 20150 &0.70 & 1.79 & 4.07 & 0.88 & 3812& 3708& 3790& 3747 &3976& 3949 & 3859 & 3801 &3830 \\
II85  & 40056 &0.70 & 1.57 & 3.85 & 0.78 & 3952& 3866& 3845& 3819 &4170& 4170 & 3989 & 3952 &3970 \\
IV13  & 20074 &0.70 & 1.54 & 3.68 & 0.78 & 3976& 3885& 3914& 3894 &4207& 4138 & 4032 & 3972 &4002 \\
III3  & 40082 &0.70 & 1.47 & 3.61 & 0.75 & 4040& 3951& 3932& 3922 &4268& 4225 & 4080 & 4033 &4056 \\
\noalign{\smallskip} \hline \end{tabular}
\end{flushleft} 
\end{table*}

\begin{figure} 
\includegraphics[width=9cm]{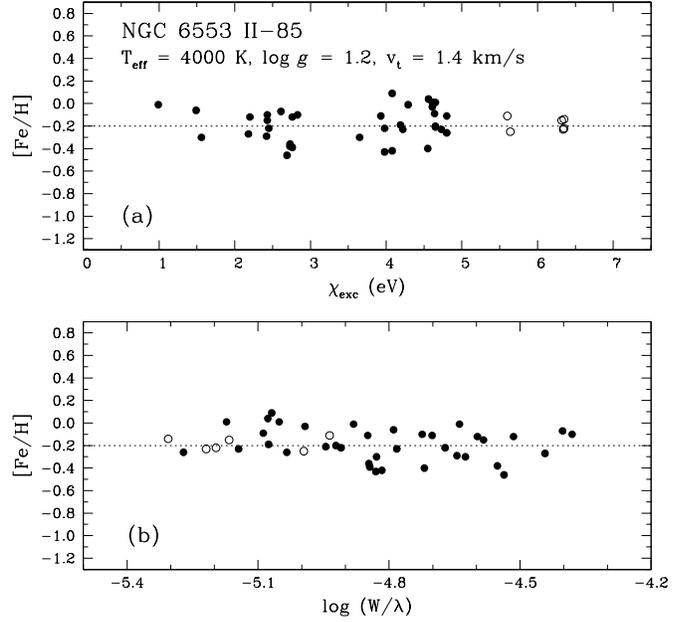}
\caption[]{Iron abundances obtained from optical (filled circles) 
and infrared (open circles) Fe I lines (for NGC6553 II-85) as a
function of (a) excitation potential and (b) reduced equivalent width. 
The dotted line is the mean abundance obtained from IR Fe I lines.
}
\label{figc}
\end{figure}

\subsection { Gravities}
 The classical relation 
$\log g_*=  4.44 + 4\log T_*/T_{\odot} + 0.4(M_{\rm bol}-M_{\rm bol\odot})
+ \log M_*/M_{\odot}$ 
was used (adopting T$_{\odot}$ = 5770 K, M$_*$ = 
0.85 M$_{\odot}$ (Bertelli et al. 1994) 
and M$_{\rm bol \odot}$ = 4.75  
(Cram 1999; see also Bessell et al. 1998).
For deriving M$_{\rm bol *}$ we adopted the
distance
modulus (m-M)$_{\circ}$ = 13.6 
and A$_{\rm K}$ = 0.38E(B-V) (Guarnieri et al. 1998).
The bolometric magnitude
corrections BC$_K$ are from  Houdashelt
et al. (2000a,b). 
The derivation of  M$_{\rm bol}$ =  M$_K$ +  BC$_K$
 is better than that using  M$_V$ + BC$_V$
because it is  much less dependent on reddening and metallicity. 
The resulting M$_{\rm bol}$  values
are given in Table 7.

\subsection{Equivalent Widths, Oscillator Strengths
and Damping Constants}

Equivalent widths of \ion{Fe}{I} lines 
were measured from the infrared spectra, and
from the  optical spectrum of the star II-85 (see Sect. 3.2).
The list of \ion{Fe}{I} lines in the H band is given in Table 4.

\begin{table*}
\caption[1]{Equivalent Widths of Infrared Fe I Lines}
\begin{flushleft}
\begin{tabular}{clllllllllrrrrrlll}
\hline \hline
\noalign{\smallskip}
${\rm \lambda}$ & ${\rm \chi_{exc}}$ & 
\multicolumn{7}{c}{log $gf$} & & \multicolumn{5}{c}{W (m\AA)}\\
\cline{3-9} \cline{11-15} \\
    {(\AA)}      &     (eV)        &  adopted  & TWG$^1$ & TWP$^1$ & 
MB99$^2$ & S02$^1$ & VMB95$^2$ & MS92$^2$ & & 40201 & 20150 & 40056 & 20074 & 40082 \\
\noalign{\smallskip}
\hline
\noalign{\smallskip}
15531.753$^a$ &5.64& -0.48 & -0.47& -0.49& $^a$     &-0.56    & -0.45$^e$& ---     &&177 &180& 157&160 &179\\
15534.260    & 5.60& -0.45 & -0.47& -0.50& -0.47    &-0.40    & -0.28$^e$&-0.34$^i$&&194 &180& 180&190 &156\\
15537.690$^b$ &6.32& -0.23 & -0.26& -0.21&  $^b$    &-0.80$^d$& -0.21$^g$& ---     &&116 &109& 106&106 &107\\
15550.450$^c$ &6.34& -0.25 & -0.27& -0.22& -0.35$^c$&  ---    & -0.23$^g$& ---     && 84 &110&  94& 83 & 96\\
15551.430 & 6.35& -0.20 & -0.21   & -0.16& -0.31    &  ---    & -0.16$^h$& ---     && 80 &100&  99& 76 &105\\
15566.725 & 6.35& -0.45 & -0.50   & -0.45& -0.50    &  ---    & -0.36$^h$&-0.43    && 57 & 71&  77& 66 & 71\\
\noalign{\smallskip} \hline
\end{tabular}
\renewcommand{\arraystretch}{1}
\begin{list}{}{}
\item [$^1$] based on Arcturus spectrum
\item [$^2$] based on Solar spectrum
\item [$^a$] blend (5.64 eV \& 6.24 eV)
\item [$^b$] blend (6.32 eV \& 5.79 eV)
\item [$^c$] blend (6.32 eV \& 6.36 eV)
\item [$^{d-i}$] The excitation potential adopted was: 
$\chi_{exc}$ = 5.79, 5.71, 6.39, 6.42 \& 5.64 eV, respectively.
\item References: TWG: This work (Gustafsson model); TWP: This work (Plez model); 
MB99; S02: Smith et al. (2002);
VMB: Valenti, Marcy \& Basri (1995); MS92: Muglach \& Solanki (1992)
\end{list}
\end{flushleft} 
\end{table*}

Fig. 6b shows  the derived Fe abundances as a function
of reduced equivalent widths.
The equivalent widths in the optical were measured using
DAOSPEC,  a new  automatic code
developed  by  P.Stetson (Stetson  et  al.    2003,  in
preparation). 
In Barbuy et al. (1999), with  a  lower S/N and  resolution
(R = 20 000), only 
 stronger lines were measured, and
 a low metallicity derived in Barbuy et al. (1999) 
of [Fe/H] = -0.55 was probably underestimated. 
In the present optical and IR spectra, of higher resolution, 
the deblending with CN lines is also much improved:
the Arcturus spectrum (Hinkle et al. 1995) was inspected 
in order to avoid blends with CN lines.

\begin{figure*} 
\includegraphics[width=18cm]{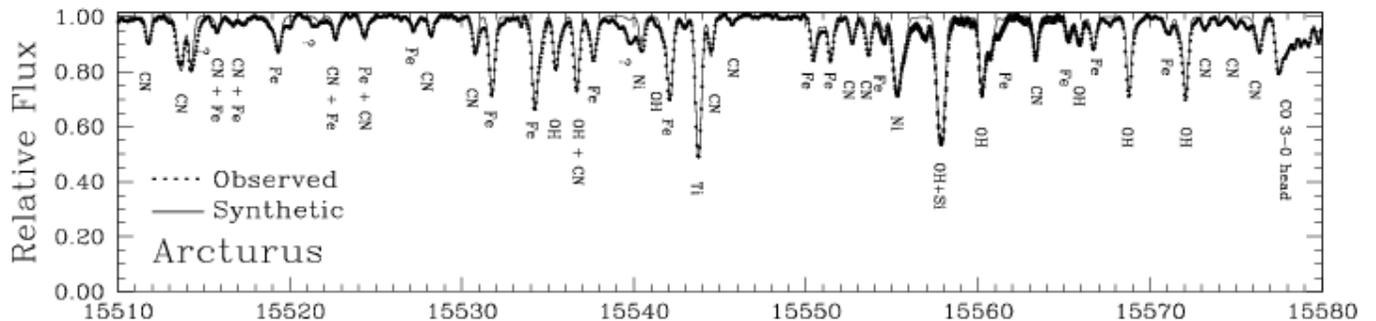}
\caption[]{Observed (dotted line) and synthetic (solid line) 
spectra of Arcturus in the region 1.551-1.558 $\mu$m.
}
\label{figc}
\end{figure*}

\begin{table}
\caption[6]{Parameters and abundances for Arcturus}
\begin{flushleft}
\begin{tabular}{lllllllllllll}
\noalign{\smallskip}
\hline
\noalign{\smallskip}
{\rm Parameter} & This Work & Literature$^*$ \\
\noalign{\smallskip}
\hline
\noalign{\smallskip}
${\rm T_{eff}}$ (K)& 4275 & 4300 $\pm$ 10 \\
log $g$       & 1.55  & 1.55 $\pm$ 0.10 \\
mass (M$_{\odot}$)  & 0.9  & 0.81$^{**}$ $\pm$ 0.20 \\
${\rm v_t (km s^{-1})}$ & 1.65 & 1.67 $\pm$ 0.05 \\
{[Fe/H]}          & -0.54 & -0.54 $\pm$ 0.06 \\
{[C/Fe]}          & -0.08 & -0.04 $\pm$ 0.04 \\
{[N/Fe]}        &  +0.30 &  +0.27 $\pm$ 0.08 \\
{[O/Fe]}        &  +0.43 &  +0.39 $\pm$ 0.06 \\
{[Ni/Fe]}       &  +0.02 &  +0.05 $\pm$ 0.05 \\
\noalign{\smallskip} \hline \end{tabular}
\renewcommand{\arraystretch}{1}
\begin{list}{}{}
\item [$^{*}$] References: Brown \& Wallerstein (1992);
Peterson et al. (1993); McWilliam \& Rich (1994); 
Sneden et al. (1994); Balachandran \& Carney (1996);
Hill (1997); Gonzalez \& Wallerstein (1998);
Thevenin \& Idiart (1999); Tomkin \& Lambert (1999);
Griffin \& Lynas-Gray (1999);
Smith et al. (2000); Carr et al. (2000);
Decin et al. (2000); Mishenina \& Kovtyukh (2001);
Smith et al. (2002); Reddy et al. (2002).
\item [$^{**}$] obtained from log $g$ (= 1.55 $\pm$ 0.10) 
and R (= 25R$_{\odot}$)
\end{list}
\end{flushleft} 
\end{table}

\subsection{Spectrum synthesis}

We have adopted the photospheric models for giants of Plez et al. 
(1992).

The atomic parameters derived in Mel\'endez \& Barbuy (1999, hereafter MB99)
were revised based on the spectrum of Arcturus,
using the Plez et al. (1992) model atmosphere,
 in order to be consistent with the
models used for the sample stars.

In Table 4 are reported the stellar parameters
found in the literature for Arcturus and the 
presently adopted parameters obtained in the same
way as for stars of NGC 6553.
A fit to the spectrum of Arcturus
(Hinkle et al. 1995) in the H band
is shown in Fig. 7. The metallicity
[Fe/H] = -0.54  adopted here
is rather different from the value adopted
in Smith et al. (2002) of [Fe/H] = -0.72,
but the [O/Fe] = +0.43 (present work) 
and +0.34 (Smith et al. 2002) values are in close
agreement.  

The revised
 oscillator strengths of Fe I lines are given in Table 4.
 The damping constants
were computed based on the  
collisional broadening theory of 
Barklem et al. (1998a), using tables by
 Anstee \& O'Mara (1995), Barklem \& O'Mara (1997) and
Barklem et al. (1998b).
Molecular lines of CN A$^2$$\Pi$-X$^2$$\Sigma$,
OH X$^2$$\Pi$ and
CO X$^1$$\Sigma^+$ are taken into account.
See more details and the full list of lines in
MB99, Mel\'endez et al. (2001) and Mel\'endez \&
Barbuy (2002).
A difference relative to these papers is the adoption
of a dissociation potential of CN 
D$_{\circ}$(CN) = 7.72 eV (Pradhan et al. 1994),
and electronic transition moment for CN from
Bauschlicher et al. (1988).

For the optical spectrum of II-85 the Fe I gf-values
were adopted from the National Institute of  Standards \& Technology (NIST)
database  (Martin   et al. 2002).

\begin{figure*} 
\includegraphics[width=17cm]{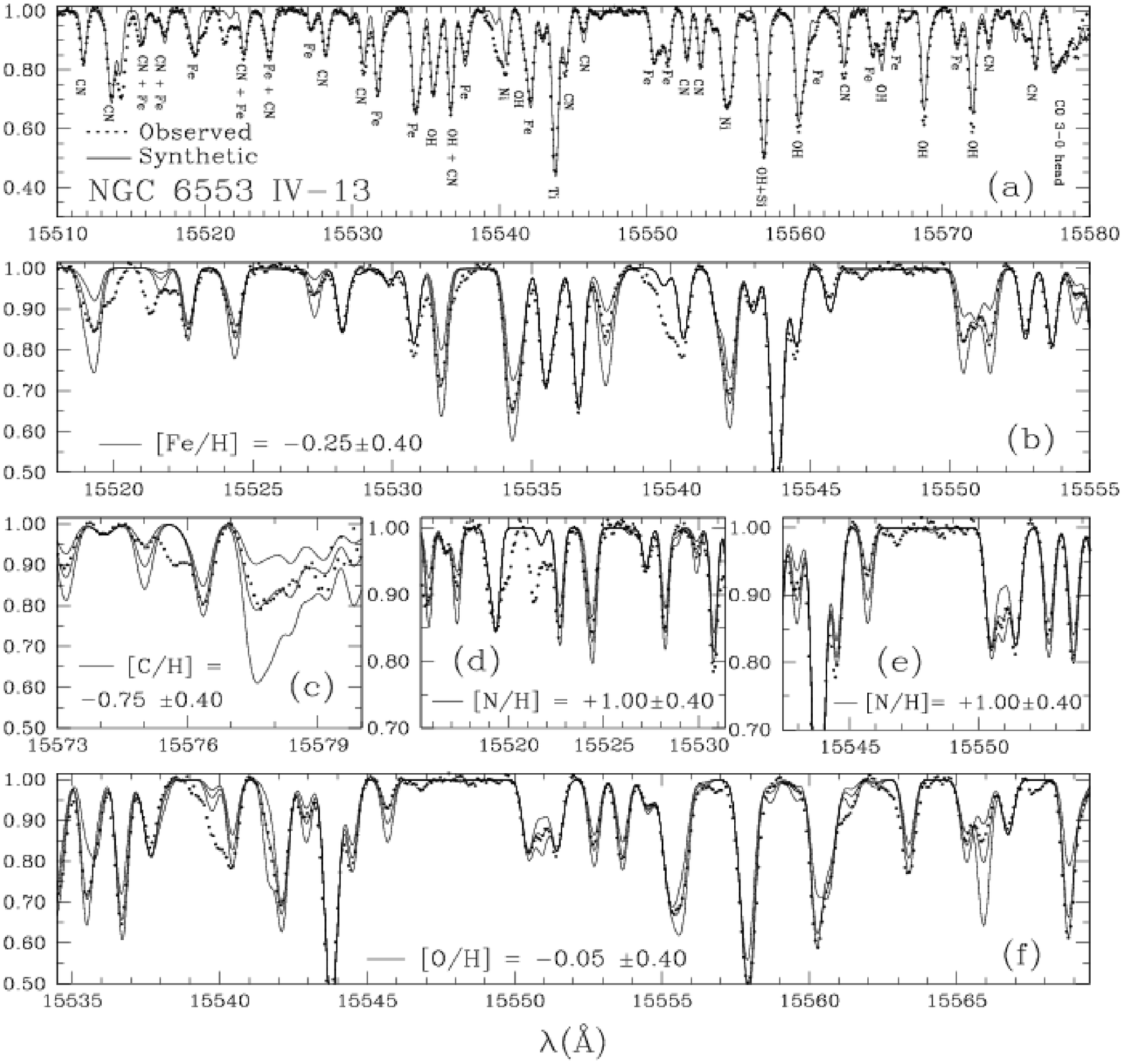}
\caption[]{Observed (points) and synthetic (line) spectra 
in the giant NGC 6553 IV-13. (a) Fit to the overall region and
identifications of atomic and molecular lines; (b)-(f) Synthetic spectra
with differing ($\pm$ 0.4 dex) Fe, C, N and O abundances, showing the 
sensitivy of iron, CO, CN and OH lines, respectively.
}
\label{fig8}
\end{figure*}

\subsection{Microturbulence velocity}

For deriving the microturbulence velocity  v$_{\rm t}$, lines
of different equivalent widths are needed. In the
H band most lines, of high excitation potential, are weak.
In order to better derive  v$_{\rm t}$, we used optical lines of
the star II-85 (Sect. 3.2)
for which we obtain  v$_{\rm t}$ =
1.4 km s$^{-1}$. We adopt this same value for
the other stars of same temperature and 
v$_{\rm t}$ =
1.5 km s$^{-1}$ for the two cooler ones. 
The microturbulence velocity might be somewhat lower,
given that from VLT-UVES data for
 3 giants of NGC 6528, with effective temperatures
in the range 4200 - 4800 K, Zoccali et al. (2003c)
determined  microturbulence velocities of
v$_{\rm t}$ = 1.1 - 1.5 km s$^{-1}$.

The adopted stellar parameters are given
in Table 7.

\subsection{Errors}

Errors due to change in stellar parameters are shown in 
Table 6, for an increase of 100 K in the effective temperature,
+0.3 dex in log $g$, and +0.3 km s$^{-1}$ in v$_{\rm t}$.

\section{Results}

We obtained a mean metallicity [Fe/H] = -0.20$\pm$0.10.
 In Table 7 are reported the abundance ratios obtained
for C, N and O.
 Fig. 8 shows the fit of synthetic spectra
to the observed spectrum of star IV-13.

The C abundance determination is based on the CO 3-0 band
at 1.5578 $\mu$m. N is from a large number of CN lines
as can be seen in Fig. 8.
The analysed  giants clearly show a C deficiency and N enhancement,
which are signatures of convective mixing.
Oxygen on the other hand is little affected by the mixing,
and should reflect the original oxygen abundance.
Note that log(C+N) = 9.1, whereas the solar value is
log(C+N)$_{\odot}$ = 8.65. Given that we do not expect
that the first dredge-up modifies the C+N abundance,
this would point out to an original enhancement of
C+N in NGC 6553.   A Nitrogen original enhancement
seems likely. For such studies, in the future it will
be interesting to observe stars in the main sequence
of this cluster.

 The O abundance is derived from OH lines
of moderate strength, particularly those at 1.5536 
and 1.5566 $\mu$m.
The mean oxygen-to-iron abundance is [O/Fe] = 
+0.20$\pm$0.10.
The enhancement of oxygen points to an important contribution
from supernovae type II in the original chemical
 enrichment of this cluster.

\subsection{Comparison with previous work}

 Barbuy et al. (1999) derived [Fe/H] =
-0.55, together with  $\alpha$-element enhancements 
of [Mg/Fe] = +0.33, [Si/Fe] = +0.35,  [Ca/Fe] = +0.32
and   [Ti/Fe] = +0.50,
for two giants of effective temperatures around 4000 K.

Cohen et al. (1999) analysed 5 HB stars
and found a mean value of  [Fe/H] = -0.16, with
the mean overabundances of $\alpha$-elements
[Mg/Fe] = +0.4, [Si/Fe] = +0.14, [Ca/Fe] = +0.26,
[Ti/Fe] = +0.19, whereas for oxygen a value of
[O/Fe] = +0.5 may be overestimated given the use
of the OI 777nm triplet (e.g. Mel\'endez et al. 2001).
There is an interesting similarity
of the abundance ratios by Cohen et al.'s and
 the results
for Baade's Window field
stars by McWilliam \& Rich (1994, hereafter MR94)
where [Mg/Fe] $\approx$ [Ti/Fe]
$\approx$ +0.3 and [Ca/Fe] $\approx$ [Si/Fe] $\approx$ 0.0,
to be noted 
in particular the lower overabundance of Si and Ca, relative to 
that of Mg. In MR94 Ti is also more overabundant than Si and Ca.

Carretta et al. (2001) analysed 4 HB stars of NGC 6528,
and obtained [Fe/H] = +0.07. 
For NGC 6553, they
 proposed a rescaling of metallicity
by summing up spectra of 3 HB stars of NGC 6553
and 2 stars of NGC 6528. By comparing the
equivalent widths of the hypothetical
average red HB stars, they
rederive the stellar parameters and find a higher metallicity
compared to that found by Cohen et al. (1999).
They adopt a final mean metallicity of [Fe/H] = -0.06.
Note that, as concerns the metallicity [Fe/H],
the present results are in better agreement with Cohen et 
al (1999) than with the rescaling of Carretta et al. (2001).

A good agreement between the present results and those
by Origlia et al. (2002) is found. Their two cool giants of effective
temperature T$_{\rm eff}$ = 4000 K, analysed from
Keck-NIRSPEC infrared spectra
in the H band, at a resolution of R $\approx$ 25 000,
gave a  metallicity of 
[Fe/H] = -0.3 and [O/Fe] = +0.3; for  other
$\alpha$-elements the authors indicate  [$\alpha$/Fe] = +0.3,
but no detailed elemental abundances are given.

The main causes of disagreement with previous work
come from temperature  and microturbulence velocity
 uncertainties, and
from the resolution and S/N of the spectra.

\begin{table}
\caption[6]{Sensitivity to Stellar Parameters}
\begin{flushleft}
\begin{tabular}{llll@{}ll}
\noalign{\smallskip}
\hline
\noalign{\smallskip}
{Element} & $\Delta$ ${\rm T_{eff}}$ & {$\Delta$ log $g$} & 
{$\Delta$ v$_t$} & {$\sqrt{\Sigma x^2}$} \\
          & {+100 K} & {+0.3 dex} & {+0.3 km s$^{-1}$} & \\
\noalign{\smallskip}
\hline
\noalign{\smallskip}
{Fe [Fe I optical]}  & -0.05 & +0.07 & -0.13 & 0.16 \\
{Fe [Fe I (6 eV)]}  & -0.01 & +0.04 & -0.05 & 0.06 \\
{C  [CO]}     & +0.05 & +0.10 & -0.03 & 0.12 \\
{N  [CN]} &-0.20& -0.09 & -0.03 & 0.22 \\
{O  [OH]}           & +0.22 & +0.20 & +0.02 & 0.30 \\
{Ni [Ni I]}                  & -0.02 & +0.06 & -0.06 & 0.09 \\
\noalign{\smallskip} \hline \end{tabular}
\end{flushleft} 
\end{table}

\begin{table*}
\caption[1]{Adopted stellar parameters and abundances 
for the five sample red giants in NGC6553}
\begin{flushleft}
\begin{tabular}{lllllllllllll}
\hline \hline
\noalign{\smallskip}
${\rm Star}$ & ${\rm T_{eff}}$ & {\rm log $g$} & {\rm v\tiny{t}} &
{\rm M$_{bol}$} & [Fe/H] & [C/Fe] & [N/Fe] & [O/Fe]  & [Ni/Fe]  \\
             &     (K)        &      & (${\rm km s^{-1}}$)  &
(mag) &       &   &   &   &       \\
\noalign{\smallskip}
\hline
\noalign{\smallskip}
40201 (III-17)& 3800 & 0.8 & 1.5 & -2.2& -0.25& -0.60 &+1.20 &+0.30  &-0.10 \\
20150         & 3800 & 1.0 & 1.5 & -1.8& -0.16& -0.75 &+1.45 &+0.15 &-0.15 \\
40056 (II-85) & 4000 & 1.2 & 1.4 & -1.6& -0.20& -0.70 &+1.30 &+0.25  &-0.15 \\
20074 (IV-13) & 4000 & 1.3 & 1.4 & -1.4& -0.25& -0.50 &+1.25 &+0.20  &-0.10 \\
40082 (III-3) & 4000 & 1.4 & 1.4 & -1.1& -0.17& -0.55 &+1.30 &+0.15  &-0.10 \\
\noalign{\smallskip}
\hline
\noalign{\smallskip}
 median &   &   &     &     & -0.20& -0.60 &+1.30 &+0.20  &-0.10 \\
\noalign{\smallskip} \hline
\end{tabular}
\renewcommand{\arraystretch}{1}
\begin{list}{}{}
\item NOTE : A$_{\odot}$(Fe, C, N, O, Ni) = 
7.50, 8.55, 7.97, 8.87, 6.25.
\end{list}
\end{flushleft} 
\end{table*}

\section{Discussion and conclusions}

We have determined a metallicity of [Fe/H] = -0.20$\pm$0.10 for NGC
6553.
We find an enhancement
of C+N, corresponding to about 0.5dex relative
to the solar value, which indicates an excess of C+N 
abundance in the gas from which NGC 6553 formed.
The $\alpha$-element oxygen is 
enhanced    by [O/Fe] $\approx$ +0.20. Cohen et al. (1999)
found enhancements of other $\alpha$-elements,
with [Mg/Fe] = +0.4, [Si/Fe] = +0.14, [Ca/Fe] = +0.26,
and [Ti/Fe] = +0.19, and a mean of [$\alpha$/Fe] = 0.25.
These results altogether point to a clear enhancement of
$\alpha$ elements in stars of NGC 6553.
 
Matteucci \& Brocato (1990) and Matteucci et al. (1999) have
computed models  of bulge formation with a fast star formation
rate, which is
completed for the bulk of the bulge, in a timescale shorter than
that of SNIa explosions.
The predictions by Matteucci et al. (1999) indicate
higher [$\alpha$/Fe] values than found here, but in any case
the enhancements found here are basically consistent with  
that
chemical evolution model,
 and with the old age of NGC 6553 determined by
Ortolani et al. (1995).

Taking into account the overabundance of
 [O/Fe] = +0.20$\pm$0.10, together with results
by Cohen et al. (1999), we find an overall
enhancement of [$\alpha$-elements/Fe]
= +0.2, and  an overall
metallicity [M/H] = [Z] = -0.06, or essentially solar.

These results indicate that 
 NGC 6553 was enriched predominantly by Type II
SNs, and that the bulge underwent fast
star formation as modeled by Matteucci et al. (1999),
and is consistent with the old age derived for this cluster
by Ortolani et al. (1995).

\begin{acknowledgements} This work is
based on observations obtained at the Gemini Observatory, which is operated by the Association
 of Universities for Research in Astronomy, Inc.,
      under a cooperative agreement with the NSF on behalf of the Gemini partnership: 
the National Science Foundation (United States), the Particle
      Physics and Astronomy Research Council (United Kingdom),
 the National Research Council (Canada), CONICYT (Chile), the Australian Research
      Council (Australia), CNPq (Brazil), and CONICET (Argentina).
We acknowledge partial financial support from  CNPq
and Fapesp (Brazil).
\end{acknowledgements}

\end{document}